\documentclass[twocolumn,twoside]{article}
\usepackage{graphics}
\newcommand{\h}{\linebreak \hspace*{3ex}}
\newcommand{\hb}{\\ \hspace*{2ex}}

\begin{document}
\title{ABOUT COMPILED CATALOGUE OF SPECTROSCOPICALLY DETERMINED $\alpha$-ELEMENTS ABUNDANCES FOR STARS WITH ACCURATE PARALLAXES}
\author{T.V.\,Borkova$^{1}$, M.S.\,Katchieva$^{1}$, B.A.\,Marsakov$^{1}$, 
D.M.\,Pitkina$^{1}$, \\[2mm] 
\begin{tabular}{l}
 $^1$ Southern Federal University\hb
 Rostov-on-Don 344090 Russia, \\
 {\em borkova@ip.rsu.ru}\\
\end{tabular}
}
\date{}
\maketitle

ABSTRACT.
We present a new version of the compiled catalogue of nearby 
stars for which was published the spectoscopically determined 
effective temperatures, surface gravities, and abundances of 
iron, magnesium, calcium, silicon, and titanium. Distances, 
velocity components, galactic orbital elements, and ages was 
calculated for all stars. The atmospheric parameters and iron 
abundances were found from 4700~values in 136~publications, 
while relative abundances of alpha-elements were found from 
2800~values in 81~publications for $\approx 2000$~dwarfs and 
giants using a three-step iteration averaging procedure, with 
weights assigned to each source of data as well as to each 
individual determination and taking into account systematic 
deviations of each scale relative to the reduced mean scale. 
The estimated assumed completeness for data sources containing 
more than five stars, up to late April~2007, exceeds 90\,\%. 
For the vast majority of stars in the catalogue, the 
spatial-velocity components were derived from modern 
high-precision astrometric observations, and their Galactic 
orbit elements were computed using a three-component model 
of the Galaxy, consisting of a disk, a bulge, and a massive 
extended halo. Ages was determined for dwarfs and subgiants 
using Yale isochrones~2004. For this purpose the original 
codes was developed, based on interpolation with the 3D-spline 
functions of theoretical isochrones, and with subsequent 
interpolation in  metallicity and abundances of $\alpha$-elements.
\\[1mm]

{\bf Galaxy (Milky Way), stellar chemical composition, thin disk, 
Galactic evolution.}\\[2mm]

\begin{figure}
\resizebox{\hsize}{!}
{\includegraphics{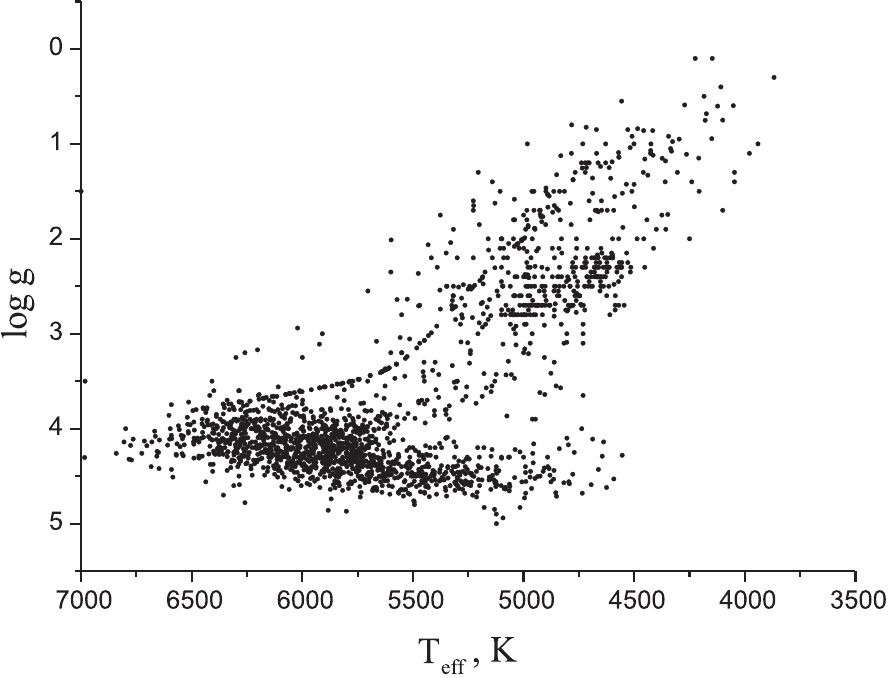}}
\caption{The  $T_{\textrm{eff}}-\log g$ diagram for stars with 
$\alpha$-element abundances found in the literature.}
\label{fig_1}
\end{figure}

The various published abundances of an element for a 
given star often differ quite appreciably, even when the 
spectra reduced by different authors are of similarly high 
quality. If several abundance values are 
available for the same star, they can simply be averaged. However, when 
an abundance is presented in only one paper, the possibility of 
systematic differences must be considered. We collected all available 
lists (with $\geq$\,5\,stars) of relative abundance estimates of four 
$\alpha$-elements ([Mg/Fe], [Ca/Fe], [Si/Fe], [Ti/Fe]) for field 
stars from high-resolution spectra with high signal-to-noise ratios 
published after 1989. We estimate the completeness of the abundances 
published for solar-vicinity stars up through April~2007 to be 
better than 90\,\%. The raw material for this study were 
81~publications containing 2800~$\alpha$-element-abundance 
determinations for $\approx 2000$~stars. 
The Hertzsprung-Russell 
diagram in Fig.\,\ref{fig_1} demonstrate the stars of our catalog. 

\begin{table*}
\begin{center}
\caption{Internal accuracy of final atmospheric parameters and relative 
abundances of $\alpha$-elements for catalogue stars}
\begin{tabular}{lccccccc}
\hline
[Fe/H]& $T_{eff}$& $\log g$ & $\varepsilon$[Fe/H]& $\varepsilon$[Mg/Fe] & 
 $\varepsilon$[Ca/Fe] & $\varepsilon$[Si/Fe] & $\varepsilon$[Ti/Fe]\\
range & K$^\circ$ &  & dex & dex & dex & dex & dex\\
\hline
$> - 1.0$ &  58 & 0.12 & 0.06 & 0.07 & 0.07 & 0.05 & 0.15\\
$< - 1.0$ & 137 & 0.24 & 0.09 & 0.09 & 0.09 & 0.11 & 0.15\\
\hline
\end{tabular}
\end{center}
\end{table*}

To derive reliable atmospheric parameters and abundances, 
we applied three-step iterative technique for compiling data, 
which in detail described in the paper (Borkova, Marsakov,~2005), 
with awarding of weight both to each source and to each 
determination of the averaged value. Stellar effective 
temperatures, metallicities, and relative abundances of 
$\alpha$-elements were leaded to scales of Edvardsson et al.~(1993). 
The surface gravities was leaded to scale of Gratton et al.~(2003), 
where they was determined on the basis of trigonometric 
parallaxes. We found it necessary to differentiate between 
the two metallicity groups because the uncertainties in all 
the parameters are considerably larger for the metal-poor stars.

The first step of averaging procedure was a simple mean.  
Our analysis shows that the scatter of the deviations and the 
systematic offset of individual atmospheric parameters and abundance 
determinations relative to the calculated mean values vary from 
list to list and also depend on metallicity. To take these small 
but systematic trends into account, we divided each list into 
two metallicity ranges at [Fe/H]\,$=-1.0$ and calculated the 
mean deviations for these ranges. We then corrected all the 
individual determinations of each parameter for these biases. 
These corrections leave the determined parameter for stars 
present in several lists virtually unchanged. However, if a 
star's parameter was determined in a single study only, the 
correction will strongly affect the final determined parameter.

The next step after correcting for systematic biases was to 
determine weights for the data sources and calculate new weighted means.
Each source was assigned a weight that was inversely proportional 
to the corresponding dispersion for the deviations in each 
of the metallicity ranges. In this case one and the same 
source could obtain different weight for each determined 
parameter. The lowest scatter for the higher 
metallicity range was found for the lists of Mashonkina et al.~(2003), 
Edvardsson et al.~(1993), and Jehin et al.~(1999), and they were 
assigned unit weights. At lower metallicity range, the lowest
scatter was shown by the lists of Nissen \& Schuster~(1997),
Mashonkina et al.~(2003) and Jehin et al.~(1999). The 
lowest weights assigned to some of the lists were $\approx 0.2$. 
We then calculated a new weighted mean each parameter for each 
star taking into account the biases and weights assigned to the lists.

The next step was also a weight-assigning procedure, this time 
for individual parameter determinations. This procedure 
was intended to assign lower weights to initial values showing 
larger deviations. Clearly, such a procedure can work only if 
there are three or more values for the same star. When assigning 
the weights, we considered the mean absolute value of the 
deviations for all stars in the list containing the given 
value. As a result, this procedure assigns the lowest weights 
to the least-reliable determinations and enables us to obtain 
final values that are close to those given for most of the 
sources, with no single measurement rejected.

For all parameters, we estimated the uncertainties of the 
averages based on the scatter of the individual values about 
the final average for each star; i.\,e., from the agreement of 
the values obtained by the various authors. The corresponding 
uncertainties are presented in Table~1. All these 
estimates are close to the lower limits of the 
uncertainties for these parameters claimed by the authors.

We determined the distances to the stars using 
trigonometric parallaxes with uncertainties below 20\,\%. In 
their absence we adopted the photometric distances, derived 
using uvby$\beta$ photometric data. The uncertainty in photometric 
distances is usually claimed to be $\pm 13$\,\%.

We took the proper motions from the catalogs Hipparcos~(1997), 
in their absence we adopted other background catalogs. 
spatial velocities and galactic orbital elements we computed the 
$U$, $V$, and $W$ components of the total spatial velocity 
relative to the Sun for stars with distances, proper motions,
and radial velocities. The main contribution to
the uncertainties in the spatial velocities comes from the 
uncertainties in the distances, rather than the uncertainties 
in the tangential and radial velocities. For mean distance 
uncertainties of 15\,\% and the mean distance from the Sun of 
the sample stars, $\approx 60$\,pc, the mean uncertainty in 
the spatial velocity components is $\approx \pm 2$\,km/s.

We calculated the Galactic orbital elements by modeling 30~orbits 
of each star around the Galactic center using the multi-component 
model for the Galaxy of Allen \& Santillan~(1991), which consists 
of a disk, bulge, and extended massive halo.

Ages were determined on the basis of Yale isochrones~(2004) 
approximately for 1000~dwarfs and subgiants. For this purpose 
was developed the original procedure of 3D-spline 
interpolation of published theoretical isochrones. Procedure 
considers not only the metallicity of star, but also the content 
of $\alpha$-elements in it. 

The complete describing of the catalog will be published 
latter in Astronomical Repots.

{\it Acknowledgements.} This work was supported in part by the 
Federal Agency for Education (projects RNP~2.1.1.3483 and 
RNP~2.2.3.1.3950) and by the Southern Federal University (project 
K07T~-- 125).
\\[3mm]
\indent

{\bf References\\[2mm]}
1.C.~Allen and A.~Santillan, Rev. Mex. Astron. y Astrofis.
{\bf 22}, 255 (1991)\\
2.~T.V.~Borkova and V.A.~Marsakov, Astron.Zh. {\bf 82}, 453 (2005)
[Astron.Rep.49 405 (2005)].\\
3.~B.~Edvardsson, J.~Andersen, B.~Gustafsson, \emph{et al.}, Astron.
and Astrophys. {\bf 275}, 101 (1993).\\
4.~R.~G.~Gratton, E.~Carretta, S.~Desidera, \emph{et al.}, Astron.
and Astrophys. {\bf 406}, 131 (2003)\\
5.~E.~Jehin, P.~Magain, C.~Neuforge, \emph{et.al.}, Astron. and
Astrophys. {\bf 341}, 241 (1999)\\
6.~L.~Mashonkina, T.~Gehren, C.~Travaglio, and T.~Borkova, Astron.
and Astrophys. {\bf 397}, 275 (2003)\\
7.~P.~E.~Nissen and W.~J.~Schuster, Astron. and Astrophys.
{\bf 326}, 751 (1997).\\
\\

\vfill
%

\end{document}